\documentclass[twocolumn]{aastex62}

\usepackage{graphicx}
\usepackage{amsmath,bm}
\usepackage{CJKutf8}

\usepackage{xcolor}
\usepackage[normalem]{ulem}

\submitjournal{the Astrophysical Journal Letters}

\shorttitle{PSP observation of alpha particles heating}
\shortauthors{Xi et al. 2026}

\begin{document}

%\title{Preferential Heating of Protons over Alpha Particles near Turbulent Coherent Structures: Parker Solar Probe Observations}
\title{Parker Solar Probe Observations of Preferential Heating of Protons over Alpha Particles near Turbulent Coherent Structures}

\correspondingauthor{Tieyan Wang}
\email{tieyan.wang@gmail.com}

\author[0009-0000-4238-6739]{Jiayang Xi}
\affil{School of Earth Science, Yunnan University, Yunnan 650091, China}

\author[0000-0003-3072-6139]{Tieyan Wang}
\affil{School of Earth Science, Yunnan University, Yunnan 650091, China}

\author{Daniel Verscharen}
\affiliation{Mullard Space Science Laboratory, University College London, Dorking RH5 6NT, UK}

\author{Yan Yang}
\affiliation{Department of Physics and Astronomy, University of Delaware, Newark, DE, 19716, USA}

\author{Luca Sorriso-Valvo}
\affiliation{CNR, Institute for Plasma Science and Technology (ISTP), Bari, Italy}
\affiliation{Department of Electromagnetics and Plasma Physics, School of Electrical Engineering and Computer Science, KTH - Royal Institute of Technology, Stockholm, Sweden}

\author{Xinyi Wang}
\affil{School of Earth Science, Yunnan University, Yunnan 650091, China}

\author{Wenhao Chen}
\affil{State Key Laboratory of Deep Oil and Gas, China University of Petroleum (East China), Qingdao 266580, China}
\affil{School of Geosciences, China University of Petroleum (East China), Qingdao 266580, China}

\author{Zuzheng Chen}
\affiliation{School of Space and Earth Science, Beihang University, Beijing, China}

\author{Zeren Zhima}
\affiliation{National Institute of Natural Hazards, Ministry of Emergency Management of China, Beijing, China}

\author{Chao Xiao}
\affiliation{College of Advanced Interdisciplinary Studies, National University of Defense Technology, Changsha, China}

\author{Xiangcheng Dong}
\affil{School of Earth Science, Yunnan University, Yunnan 650091, China}

\author{Jin Liu}
\affil{School of Earth Science, Yunnan University, Yunnan 650091, China}

\author{Xiang Li}
\affil{School of Earth Science, Yunnan University, Yunnan 650091, China}

\author{Guoqi Liu}
\affil{School of Earth Science, Yunnan University, Yunnan 650091, China}

\author{Naifei Gou}
\affil{School of Earth Science, Yunnan University, Yunnan 650091, China}

\author{Xiaoxiao Qin}
\affil{School of Earth Science, Yunnan University, Yunnan 650091, China}

\author{Malcolm Dunlop}
\affiliation{School of Space and Earth Science, Beihang University, Beijing, China}
\affil{RAL Space, STFC, Oxfordshire, UK}

\author{Jinbin Cao}
\affiliation{School of Space and Earth Science, Beihang University, Beijing, China}

\begin{abstract}

    Solar wind alpha particles exhibit preferential heating and acceleration relative to protons; however, their behavior in the vicinity of turbulent coherent structures remains less understood. We report the first evidence of localized alpha particle and proton heating within coherent structures identified using the Partial Variance of Increments (PVI) method, based on Parker Solar Probe (PSP) observations. Our results show that high-PVI events are associated with significant, species-dependent temperature enhancements: protons undergo a relative larger temperature increase than alpha particles. This preferential proton heating produces a localized decrease in the alpha-to-proton temperature ratio, indicating that the plasma is driven toward thermal equilibration between species. The heating is also anisotropic, being dominated by enhancements in the perpendicular temperature. These temperature-signatures coincide with a pronounced reduction in the normalized alpha–proton differential flow speed and a localized minimum in the Coulomb collision age, suggesting that the relaxation is affected primarily by collisionless kinetic effects. These findings provide new insight into the intermittent energy conversion and ion thermodynamics in the solar wind.

\end{abstract}

% \keywords{--- solar wind --- turbulence --- plasma --- kinetic processes}
% \Unified Astronomy Thesaurus concepts{--- solar wind --- turbulence --- plasma --- kinetic processes}

\section{Introduction} \label{sec:intro}

The solar wind, originating from the Sun's corona and permeating the heliosphere, is a continuous outflow of charged particles primarily comprised of protons, electrons, and a small fraction of alpha particles and heavy ions \citep{2006LRSP....3....1M}. Observations reveal that the solar wind remains hotter than predicted by adiabatic expansion, implying the presence of ongoing heating and acceleration mechanisms during its propagation through interplanetary space \citep{2003GeoRL..30.1206R}. In addition, the proton temperature exhibits anisotropy with respect to the local magnetic field, suggesting that the underlying heating processes are not isotropic but are regulated by kinetic plasma dynamics \citep{2007GeoRL..3420105M}. Elucidating the processes responsible for energy dissipation and plasma heating in the solar wind is essential for advancing our knowledge of plasma behavior in collisionless environments and for enhancing predictive models of space weather. 

The complexity of the solar wind manifests in its multiscale turbulence characteristics \citep{2019LRSP...16....5V,2021GMS...259...67E,2026ApJ...999...99T}. Turbulence mediate the nonlinear cascade of energy from macroscopic motions to microscopic scales, where dissipation processes occur \citep{1995SSRv...73....1T,2013LRSP...10....2B,2016JPlPh..82f5302C,2017PhPl...24g2306Y,2017PhRvE..95f1201Y,2023PhR..1006....1M,2023PhPl...30b0501W}. Intermittency, a hallmark of solar wind turbulence, is evidenced by the presence of coherent structures (e.g., current sheets, vortices, and shear layers) embedded within the turbulent flow \citep{2001P&SS...49.1193S,2008NPGeo..15...95A,2013PhPl...20a2303K,2016JGRA..121.3870R,2017MNRAS.466.3918M,2023MNRAS.524.5468Z,2024ApJ...971...88V,2024ApJ...971..179B,2024JPlPh..90f9002O,2025ApJ...994L..19Y}.

The organized, spatially localized concentration of intense magnetic and velocity gradients serve as preferential sites of enhanced energy transfer and dissipation \citep{2012PhRvL.109s5001W,2015RSPTA.37340154M,2016PhPl...23d2307W,2025ApJS..281...45C}. Numerical simulations and spacecraft observations provide evidence that intermittent heating frequently occurs in the vicinity of coherent structures \citep{2011ApJ...727L..11O,2012PhRvL.108z1102O,2013ApJ...763L..30W,2013ApJ...772L..14W,2015JPlPh..81a3207S,2021ApJ...921...65Y,2022ApJ...935L..29S}. Kinetic-scale analyses further reveal the distinct non-Maxwellian particle distributions, temperature anisotropies, and enhanced wave activity, indicating the presence of kinetic effects at locations of enhanced heating \citep{2012PhRvL.108d5001S,2019ApJ...871L..22W}.

The turbulent dynamics of the solar wind become considerably more complex due to its multi-component composition, including not only protons and electrons but also alpha particles and other minor ion species \citep{2016ApJ...827L...7M,2019PNAS..116..771K,2020PhRvX..10d1050K,2022ApJ...941..137R}. Among these, alpha particles play a particularly important role since, despite their relatively low number density (e.g., $n_{\alpha}/n_{p} \sim $ 0.04--0.05), they carry a significant fraction of the solar wind's mass ($\sim$ 14--17\%), momentum, ion charge density ($\sim$ 7--9\%), and energy \citep{1970JGR....75.1178R,2008PhRvL.101z1103K,2017ApJ...849..126K,2019A&A...623L...2S,2011ApJ...741...43P,2013ApJ...762...99P,2020ApJ...889L..14Z,2024ApJ...968...93X}. Observations have revealed that alpha particles in the weakly collisional fast solar wind at 1 AU are hotter than the protons \citep{1982JGR....87...52M,2011ApJ...728L...3B,2013ApJ...777L...3B,2015ApJ...812..170T,2017ApJ...849..126K}, which have been interpreted as indications of the preferential heating \citep{2019ARA&A..57..157C,2017ApJ...849..126K}. Remote observations provide additional evidence for the preferential heating of heavy ions, as their temperature ratios to protons $T_s/T_p$ often exceed the corresponding mass ratios $m_s/m_p$, where ions would attain mass proportional temperatures if their thermal speeds were equal \citep{2009LRSP....6....3C}. A variety of physical mechanisms, including the ion-cyclotron resonance \citep{2000ApJ...532.1197C,2001JGR...10624955G,2002JGRA..107.1147H,2005JGRA..110.7108G,2013PhRvL.110i1102K}, non-resonant stochastic heating \citep{2010ApJ...720..503C,2013ApJ...776...45C}, velocity filtration \citep{1992ApJ...398..299S}, and intermittent heating associated with coherent structures \citep{1999ApJ...523L..93M,2013ApJ...762...99P,2016NJPh...18l5001V}, have been proposed to account for these observations.

The Parker Solar Probe (PSP) mission \citep{2016SSRv..204....7F} has provided an unprecedented opportunity to investigate the mechanisms underlying preferential ion heating in the solar wind. Recent studies examine the near-Sun behavior and radial evolution of alpha particles \citep{2024ApJ...964L...2A,2024ApJ...963L..36L,2024ApJ...977...27P,2025ApJ...991L..35M,2026ApJ..1004...60M}. At macroscopic scales, these observations reveal that the alpha-to-proton temperature ratio $T_{\alpha}/T_p$ decreases with increasing heliocentric distance \citep{2024ApJ...963L..36L}, providing valuable insight into the spatial extent (e.g., distance from the Alfven surface) of preferential heating and the organizing influence of the local Alfvén Mach number \citep{2024ApJ...964L...2A,2024ApJ...977...27P,2025ApJ...991L..35M}. Furthermore, these works discuss the role of Coulomb collisions in regulating temperature equilibration \citep{2024ApJ...964L...2A} and the relationship between preferential heating and alpha-proton velocity drift \citep{2024ApJ...977...27P}. At kinetic scales, new insights into the fine-scale dynamics of alpha particles in different solar wind structures are investigated. \citet{2023ApJ...952L..11D} analyze the kinetic properties of alpha particles within reconnection exhaust regions, revealing signatures of localized energization and velocity-space distortions. \citet{2023ApJ...954..133H} compare the temperature characteristics of protons and alpha particles inside and outside switchback structures, showing distinct thermal responses associated with magnetic field reversals. Based on Solar Orbiter measurements \citep{2020A&A...642A...1M}, \citet{2023FrASS..1050219P} investigate the kinetic behavior of protons and alpha particles within vortex chains, identifying signatures of enhanced heating and anisotropy. More recently, \citet{2024ApJ...973..171P} examine the plasma heating near switchbacks, further supporting the idea that such coherent structures serve as efficient sites of ion energization in the inner heliosphere. The hybrid Vlasov-Maxwell simulations by \citet{2016NJPh...18l5001V} demonstrate that alpha particles experience intermittent heating similar to protons. However, different ion species can respond differently to electromagnetic fluctuations and coherent structures since their characteristic kinetic scales and resonance conditions depend on species-specific properties, including charge, mass, temperature, and drift speed. These differences may lead to distinct kinetic responses and coupling efficiencies with turbulent coherent structures. Collectively, these studies highlight the complex and species-dependent nature of the ion kinetics in the turbulent space plasma.

To the best of our knowledge, no study has systematically investigated the temperature of alpha particles near coherent structures, particularly within the inner heliospheric environment. The present study provides the first observational evidence of such processes, offering new insights into the mechanisms of species-dependent energy transfer and the role of coherent structures in the near-Sun solar wind.

\section{Data and Method} \label{sec:data}

The data used in this letter include magnetic field measurements by the FIELDS instrument \citep{2016SSRv..204...49B} and ion moments from the Solar Wind Electrons Alphas and Protons (SWEAP) instrument \citep{2016SSRv..204..131K} on PSP. For the SWEAP suite, the density, velocity, and temperature of both the proton and the alpha-particle are from the Solar Probe ANalyzer-Ions (SPAN-I) instrument, which measures 3D velocity distribution functions of ions in the inner heliosphere. As PSP moves closer to the Sun during its encounters, the increase of its orbital speed (e.g., 190 $\rm{km\,s^{-1}}$ in \citet{2022ApJ...938..138L}) causes the solar wind flow in the spacecraft frame to shift toward its ram-facing side, entering the field of view. Hence we use data from PSP's 6th to 24th encounter phases between September 2020 and June 2025, where SPAN-I is optimized for ion observations. These observations cover heliocentric distances of 0.05--0.28 AU. To ensure data integrity, we apply data quality flags to remove potentially problematic entries. In addition, both proton and alpha particle data are resampled to a 3.5-second time resolution. 

The Partial Variance of Increments (PVI) analysis method \citep{2008GeoRL..3519111G} is used to identify intermittent structures within turbulence. This technique quantifies non-Gaussian statistical properties of physical parameters, enabling the detection of characteristic coherent structures such as rotational/tangential discontinuities, current sheets, and vortices. The PVI of the magnetic field is defined as 

\begin{equation}
    \label{equ:equ1}
\mathrm{PVI}(t,\tau) = \frac{\lvert \Delta \bm{B}(t,\tau) \rvert}{\sqrt{\langle \lvert \Delta \bm{B}(t,\tau) \rvert^2 \rangle}}
\end{equation}

where $\Delta \bm{B}(t,\tau) = \bm{B}(t+\tau) - \bm{B}(t)$ is the magnetic field increment at time $t$ and scale $\tau$, $|...|$ denote the norm of the increment, and $\langle ... \rangle$ denotes a time average over the dataset. A window of 6 hours is chosen in this study, which is a few times of the turbulence correlation time in the near-Sun solar wind \citep{2020ApJS..246...53C}. High PVI values indicate the presence of strong gradients and non-Gaussian features, which are indicative of coherent structures. Events with PVI values exceeding a threshold of 3 corresponds to the upper tail of the distribution. This threshold is commonly used to identify significant intermittent structures in solar wind turbulence \citep{2011ApJ...727L..11O,2013ApJ...762...99P,2015JPlPh..81a3207S,2021ApJ...921...65Y,2022ApJ...935L..29S}. In this study, we compute the PVI index for the magnetic field data at a time lag $\tau$ of 3.5 seconds, which corresponds to the ion moments resolution. Such timescale is located near the high-frequency range of the turbulent inertial range \citep{2022RvMPP...6...41P}, where intermittent coherent structures emerge. This choice allows us to directly correlate the identified coherent structures with the ion temperature measurements. 

\section{Results} \label{sec:results}
\subsection{Event Study}

\begin{figure*}[ht!] %%%%%%%%%%%%%%%%%% FIGURE 1
    \figurenum{1}
    \centerline{\includegraphics[width=0.85\textwidth,clip=]
    {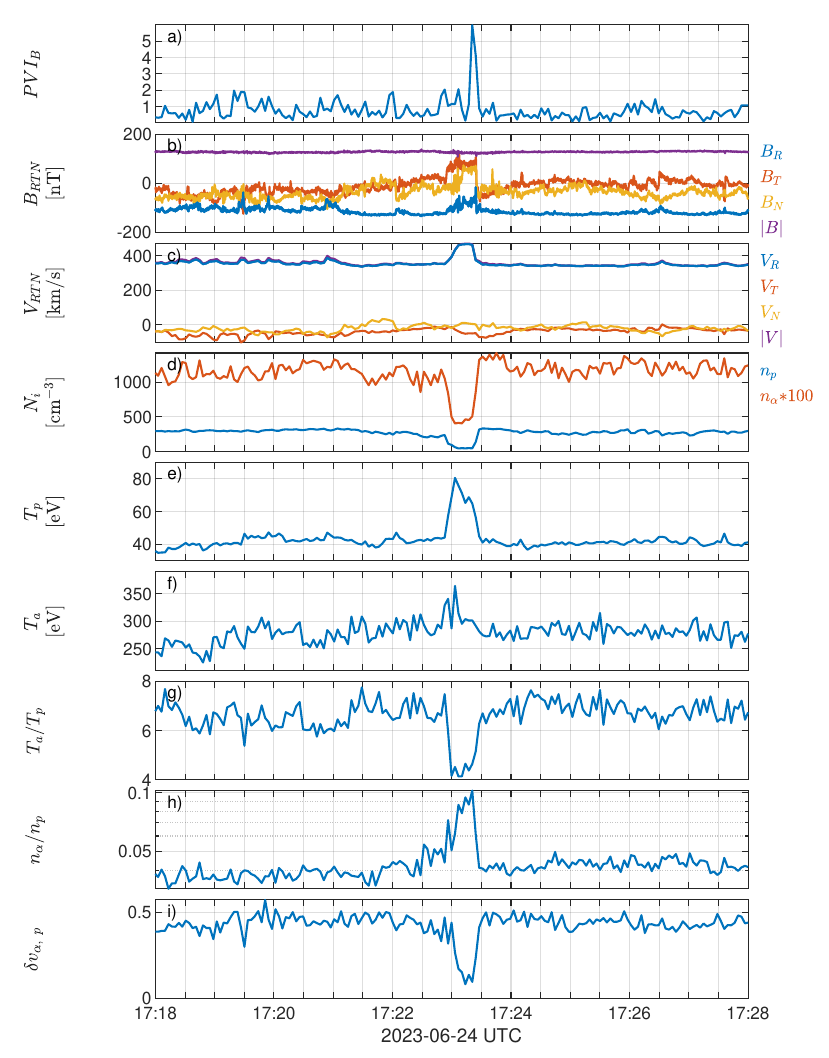}}
    \caption{Example of the differential behaviors of alpha particles and protons near the coherent structure. (a) PVI of magnetic field. (b) magnetic field $B_{RTN}$ in RTN coordinates. (c) velocity field $V_{RTN}$ in RTN coordinates. (d) number density of protons $n_p$ and alpha particles $n_\alpha$ at 100 times scale. (e) proton temperature $T_p$. (f) alpha particle temperature $T_\alpha$. (g) temperature ratio $T_\alpha/T_p$. (h) density ratio $n_\alpha/n_p$. (i) normalized drift velocity between alpha particles and protons.}
    \label{fig:fig1}
\end{figure*}

Figure \ref{fig:fig1} shows an example of the intermittent temperature enhancement on June 24, 2023. The data captures a distinct, short-duration solar wind structure characterized by a sudden spike of the PVI, reaching nearly 6 at its local maximum shown in panel a. At this 30 s event centered at approximately 17:23:15 UT, there is a noticeable change in the radial ($B_R$), tangential ($B_T$), and normal ($B_N$) components, though the total magnitude $|B|$ remains relatively constant ($\sim$ 3\% decrease) in panel b. The radial velocity ($V_R$), as the dominant component of the velocity, also displays a sharp spike, jumping from roughly 350 km/s to 500 km/s in panel c. There is a depletion in density when the velocity spikes, where $n_p$ drops from 310 $\rm cm^{-3}$ to 50 $\rm cm^{-3}$ and $n_\alpha$ drops from 11 $\rm cm^{-3}$ to 4 $\rm cm^{-3}$ in panel d. The simultaneous spike in velocity $V_R$ and the deflection in the magnetic field resembles signatures of a small-scale magnetic switchback, an S-shaped kink in the magnetic field lines with a localized jet \citep{2026SSRv..222...14B}. As shown in panel f and g, in the vicinity of the structure, the proton temperature increase from 40 eV to 70 eV and the alpha particle temperatures increase from 280 eV to a peak larger than 350 eV. The temperature ratio $T_\alpha/T_p$ drops sharply from around 7 to 4 in the event, suggesting that while both temperatures increase, $T_p$ increase more significantly relative to their base state than the alphas. In addition to the temperature signature, there is a spike in the relative abundance of alpha particles, reaching nearly 8 \%, which is larger than average solar wind values \citep{2021MNRAS.508..236W}. We also examine the normalized drift speed between two the two particle species \citep{2013PhRvL.110i1102K}, as denoted by $\delta v_{\alpha,p} \equiv |(\bm{V}_{\alpha} - \bm{V}_p)\cdot \bm{B}| / (|\bm{B}| V_A) $, where $\bm{V}_{\alpha}$ and $\bm{V}_p$ are the bulk velocities of the alpha particle and protons, respectively. $V_A$ is the local Alfvén speed calculated as $V_A = B/\sqrt{\mu_0 \left( n_p m_p + n_\alpha m_\alpha \right)}$, where $B$ is the magnetic field magnitude and $\mu_0$ is the permeability of vacuum. The drop in this quantity from around 0.5 to 0.1 during this event indicates that the relative motion between the two species is significantly reduced, and the two ion populations exhibit more similar bulk velocities. % and reduced differential streaming.

\subsection{Statistical Results of Ion Temperatures near Coherent Structures}

To gain deeper insight into the connection between the thermal signatures and coherent structures, we conduct a conditional epoch analysis of the ion temperatures around PVI-identified events \citep{2012PhRvL.108z1102O,2013ApJ...776L...8T,2018SoPh..293...10S,2020ApJS..246...46Q}. This method allows us to compare the temperature at the time of each PVI event with that of the surrounding plasma, and to examine how the temperature response depends on the strength of the intermittent structure. The analysis is performed using a total of 4,654,482 measurements at a cadence of 3.5 s. The PVI events are divided into five threshold bins: $2 \leq \mathrm{PVI} < 3$, $3 \leq \mathrm{PVI} < 4$, $4 \leq \mathrm{PVI} < 5$, $5 \leq \mathrm{PVI} < 6$, and $\mathrm{PVI} > 6$, containing 106,337, 35,913, 14,224, 6,481, and 6,186 events, respectively.

For each ion species $j=p,\alpha$, the temperature $T_j$ is evaluated as a function of the time lag $\Delta t$ relative to the reference discontinuity time $t_{\mathrm{PVI}}$. For all events whose PVI value falls within a given threshold bin $[\theta_i,\theta_{i+1})$, we collect the corresponding temperature profiles centered on $t_{\mathrm{PVI}}$ and calculate the median profile. This procedure yields a statistical estimate of the localized thermal response associated with coherent structures, while reducing the influence of individual outliers or extreme events. Formally, this can be written as:

\begin{equation}
\widetilde{T}_j(\Delta t,\theta_i,\theta_{i+1})
=
\mathrm{median}\!\left[
\frac{T_j(t_{\mathrm{PVI}}+\Delta t)}{T_{j,0}}
\;\middle|\;
\mathrm{PVI}\in[\theta_i,\theta_{i+1})
\right]
\label{eq:normalized_temp}
\end{equation}

The normalization of dividing temperature by $T_{j,0}$ allows us to isolate the localized temperature feature associated with the coherent structure from the background. The average temperature $T_{j,0}$ is calculated for each event as the mean value within an 700 s time window centered on the PVI event. The choice of the window size is motivated by the typical correlation time of solar wind turbulence at the distances studied here \citep{2020ApJS..246...53C}.

\begin{figure*}[ht!]      %%%%%%%%%%%%%%%%%% FIGURE 2
    \figurenum{2}
    \centerline{\includegraphics[width=0.95\textwidth,clip=]
    {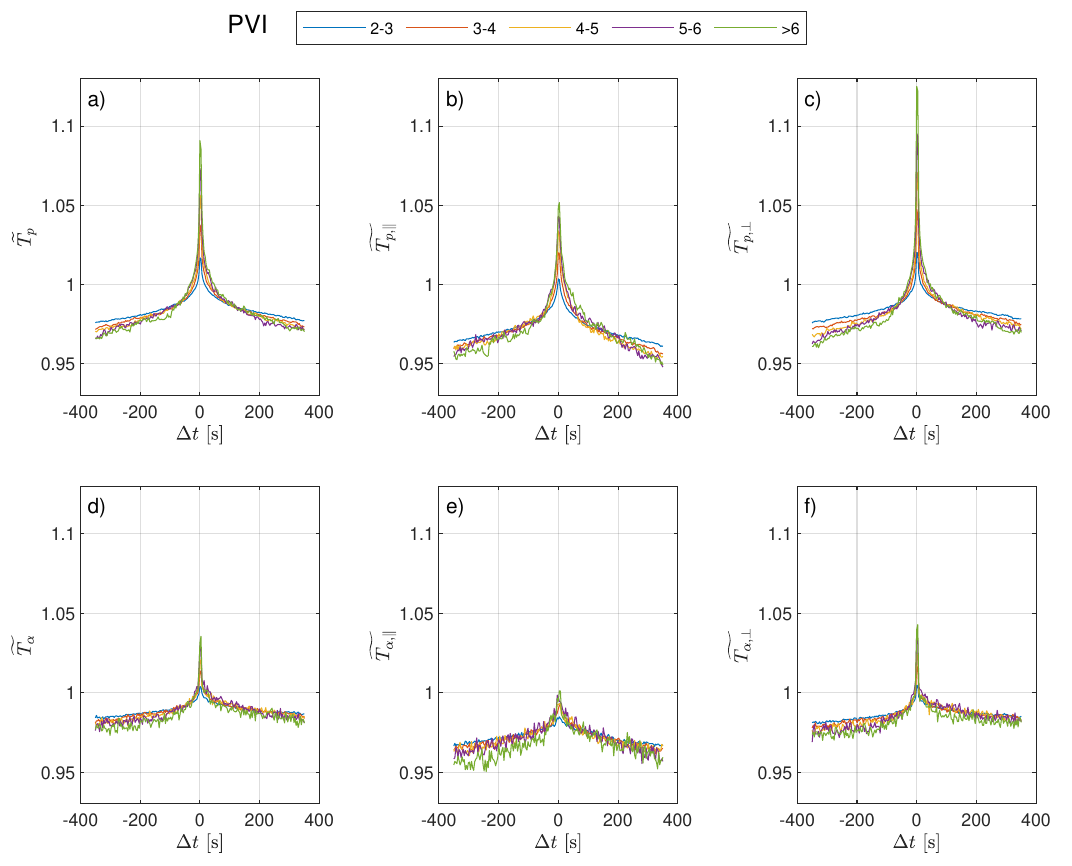}}
    \caption{Conditional median of the normalized temperature variations for protons (top row) and alpha particles (bottom row) under different PVI thresholds as defined by equation \eqref{eq:normalized_temp}. The different colored curves represent the results for 5 PVI threshold bins ($2 \leq \mathrm{PVI} < 3$, $3 \leq \mathrm{PVI} < 4$, $4 \leq \mathrm{PVI} < 5$, $5 \leq \mathrm{PVI} < 6$, and $\mathrm{PVI} > 6$). Figures 2a-2d show results for the total ($\widetilde{T_{p}}$), parallel ($\widetilde{T_{p,\parallel}}$), and perpendicular ($\widetilde{T_{p,\perp}}$) temperature changes of protons, and Figures 2d-2f show the $\widetilde{T_{\alpha}}$, $\widetilde{T_{\alpha,\parallel}}$, and $\widetilde{T_{\alpha,\perp}}$ for alphas.} %Figures 2g and 2h show the temperature anisotropy for protons and alphas, respectively.
    \label{fig:fig2}
\end{figure*}

Figure \ref{fig:fig2} shows the individual temperature response of each species. Both protons and alpha particles show a sharp increase at $\Delta t = 0$. As PVI increases (blue to green), the temperature spike becomes taller and narrower. This signature agrees with previous results \citep{2012PhRvL.108z1102O,2020ApJS..246...46Q,2022ApJ...935L..29S,2018JPlPh..84b7201S}. Despite the similar temperature features between two species, the magnitude of the temperature increase is different. The results of $\widetilde{T}_p(\Delta t,6,\rm{inf})$ in Figure \ref{fig:fig2}a show a larger spike (0.96 to 1.09, which correspond to a relative efficiency of 13\%) for the strongest PVI events (green lines) as compared to $\widetilde{T}_{\alpha}(\Delta t,6,\rm{inf})$ of alpha particles (0.97 to 1.03, which correspond to 6\% efficiency) in Figure \ref{fig:fig2}d. This suggests that the coupling efficiency between alpha particles and the turbulent structures may be weaker than protons. We investigate their temperature ratio in more details in the next section. By decomposing the temperature response into parallel and perpendicular components with respect to local magnetic field at a timescale of 3 s, the anisotropic nature of the temperatures within these intermittent regions can be highlighted. Both $\widetilde{T}_{p,\parallel}$ and $\widetilde{T}_{p,\perp}$ exhibit clear enhancements centered at $\Delta t = 0$, but the increase in the perpendicular component is more pronounced. The maximum increase in $\widetilde{T}_{p,\parallel}$ is relatively modest, reaching only approximately 8–9\% efficiency for the highest-PVI structures, whereas $\widetilde{T}_{p,\perp}$ increases by as much as 16–17\%. Moreover, the perpendicular temperature peak shows a substantially stronger dependence on PVI level than the parallel component. Despite weaker temperature increases compared to protons, the alpha particles also show a similar anisotropic response, with a parallel increase of 5\% and a more pronounced perpendicular increase of 7\% for the highest-PVI events.

\subsection{Statistical Results of Temperatures Ratios}

\begin{figure*}%[ht!]     %%%%%%%%%%%%%%%%%% FIGURE 3
    \figurenum{3}
    \centerline{\includegraphics[width=0.85\textwidth,trim ={0cm 0cm 0cm 0cm},clip=]
            {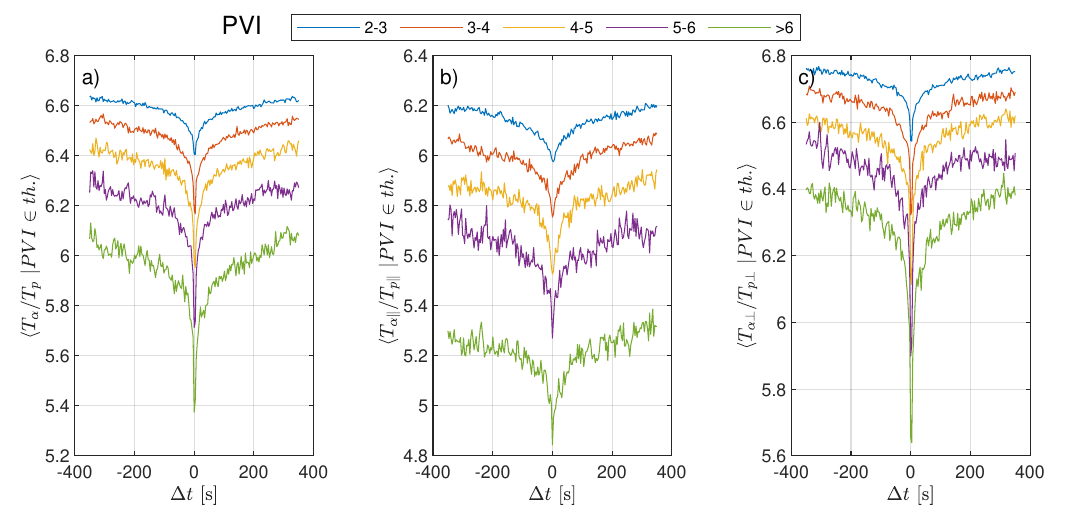}}
    \caption{Median value of the alpha to proton temperature ratio under different PVI thresholds. The columns show the total $\langle T_{\alpha}/T_{p} \rangle$, parallel $\langle T_{\alpha,\parallel}/T_{p,\parallel} \rangle$, and perpendicular $\langle T_{\alpha,\perp}/T_{p,\perp} \rangle$ temperature ratios.}
     \label{fig:fig3}
\end{figure*}

The temperature ratio $T_\alpha/T_p$ is a key diagnostic for understanding the relative thermal response of protons and alpha-particles \citep{2008PhRvL.101z1103K}. As shown in Figure \ref{fig:fig3}, sharp dips in the temperature ratio centered at $\Delta t = 0$ (the location of the PVI event) are observed in three panels. This reveals the stronger increase of protons' temperature than alpha particles, most significant near the core of the coherent structure. For the total temperature ratio, as the PVI threshold increases (moving from blue to green), the absolute value of the background ratio drops, and the depth of the dip becomes more pronounced. It reveals a clear scaling with intermittency strength. The relative decrease of $T_{\alpha}/T_{p}$ is modest, approximately 2-3\% for $2 \leq \mathrm{PVI} < 3$, but becomes progressively more pronounced with increasing PVI, approximately 4–5\% for $3 \leq \mathrm{PVI} < 4$, 5–6\% for $4 \leq \mathrm{PVI} < 5$, and nearly 12\% for the strongest events $(\mathrm{PVI} > 6)$. Since higher PVI values normally correspond to stronger or more discontinuous structures, the fact that the ratio is lower for higher PVI indicates that the strongest intermittent events in the solar wind favor higher proton temperatures over alpha-particle temperatures. The alpha-to-proton temperature ratio exhibits also a clear anisotropic response. The fractional decrease in $\langle T_{\alpha,\parallel}/T_{p,\parallel} \rangle$ remains relatively similar across different PVI levels, showing a weak dependence on PVI. In contrast, the decrease in $\langle T_{\alpha,\bot}/T_{p,\bot} \rangle$ becomes progressively stronger with increasing PVI, implying that intense coherent structures preferentially enhance proton perpendicular temperature relative to alpha particles. This demonstrates that the PVI dependence of interspecies thermal disequilibrium is primarily controlled by the perpendicular component.

\subsection{Statistical Results of Drift Speed, Collision Number, Plasma Beta, and Alpha-to-proton density ratios}

\begin{figure*}%[ht!]     %%%%%%%%%%%%%%%%%% FIGURE 4
    \figurenum{4}
    \centerline{\includegraphics[width=0.85\textwidth,trim ={0cm 0cm 0cm 0cm},clip=]
        {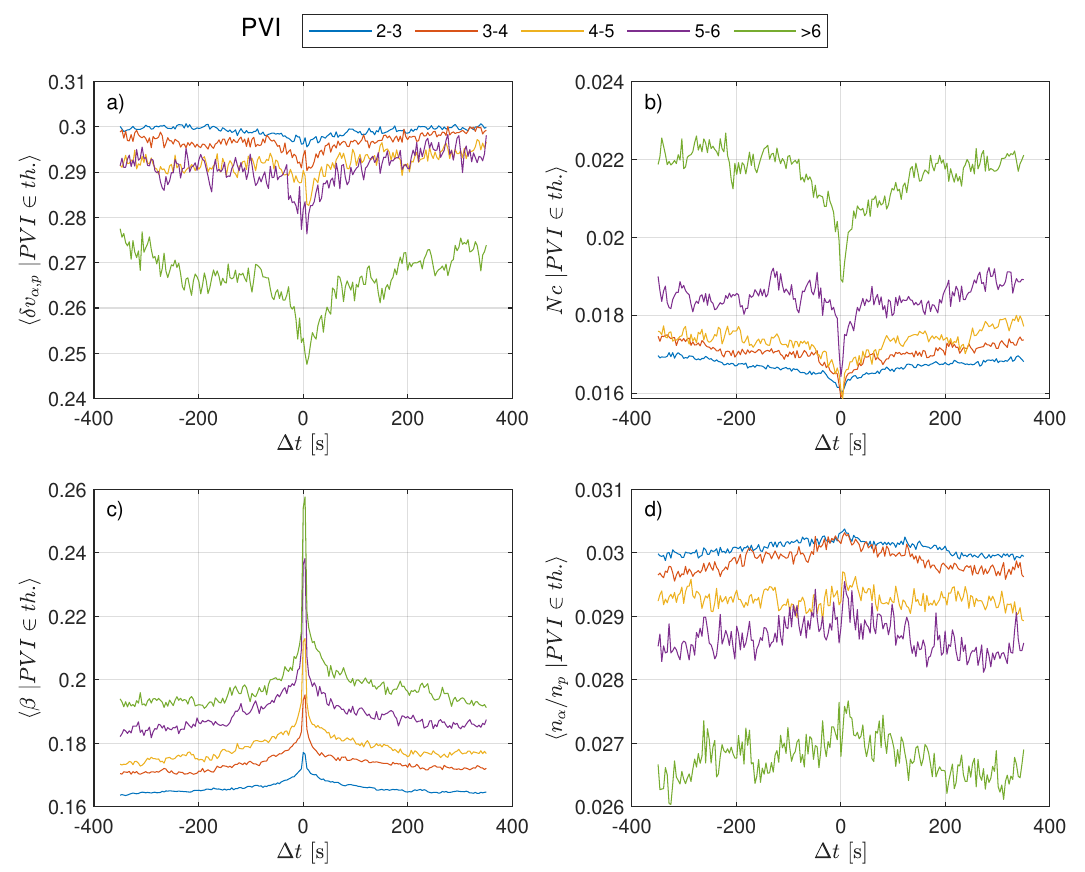}}
    \caption{
    Median of the plasma environment and composition parameters under different PVI thresholds. (a) Normalized drift speed $\delta v_{\alpha,p}$ between alpha particles and protons. (b) Collision age $N_c$. (c) Plasma beta $\beta$. (d) Alpha-to-proton density ratio $n_{\alpha}/n_p$.}
    \label{fig:fig4}
\end{figure*}

Since the near-Sun solar wind is weakly collisional and frequently exhibits notable proton–alpha differential streaming, the relative drift speed and Coulomb collision number are critical for understanding the physical processes near intermittent structures. Variations in the alpha–proton drift may indicate localized regulation of differential flow or conversion of drift energy, while the collision number provides a measure of how strongly Coulomb relaxation may have modified the non-equilibrium features. Figure \ref{fig:fig4} provides these additional context for interpreting the observed temperature features. Firstly, as shown in Figure \ref{fig:fig4}a, a collapse in the normalized drift speed $\delta v_{\alpha,p}$ is found at $\Delta t = 0$, indicating that within the core of the coherent structure, the two species are moving together more closely. The Alfvén speed $V_A$ does not change much at the event center, so the decrease in $\delta v_{\alpha,p}$ is primarily driven by the decrease in the relative velocity between the two species. Secondly, we investigate the effects of collision in Figure \ref{fig:fig4}b. The Coulomb collision number \citep{1987JGR....92.7723H,2024ApJ...964L...2A} is defined as

\begin{equation}
    \label{equ:equ3}
N_c = \frac{R}{2 V_{SW} \tau_c}
\end{equation}

where $R$ is heliographic distance, $V_{SW}$ is solar wind speed, and $\tau_c$ is the characteristic collision time for a given species, which can be calculated using the formula:

\begin{equation}
    \label{equ:equ4}
\tau_c^{-1} = \frac{32\sqrt{\pi}}{3}\ln\Lambda \,
\frac{e_\alpha^2 e_p^2}{m_\alpha m_p}
\frac{n_p}{w_{\alpha p}^3}
\end{equation}

\begin{equation}
    \label{equ:equ5}
w_{\alpha p} =
\left(
\frac{2 k_B T_\alpha}{m_\alpha}
+
\frac{2 k_B T_p}{m_p}
\right)^{1/2}
\end{equation}

where $\ln\Lambda$ is the Coulomb logarithm. $e_{\alpha}$, $e_{p}$, $m_{\alpha}$, $m_{p}$, $T_{\alpha}$, $T_{p}$ are the charges, masses, and temperatures of alpha particles and protons, respectively. $k_B$ is the Boltzmann constant and $w_{\alpha p}$ is the effective thermal speed. 

The localized dip in $N_c$ at the event center suggests that the thermal equalization may not be caused by collisional relaxation but rather driven through collisionless kinetic processes. Thirdly, the plasma beta $\beta$ as defined by the ratio of thermal pressure to magnetic pressure, displays sharp spikes at $\Delta t = 0$ in Figure \ref{fig:fig4}c. This increase is due to the increase of the thermal pressure (not shown for brevity). Lastly, to examine whether coherent structures are associated with changes in plasma composition, we plot the alpha-to-proton number density ratio in Figure \ref{fig:fig4}d. $n_{\alpha}/n_p$ remains almost constant near the centers of high-PVI structures. Examination of the individual ion densities shows that $n_\alpha$ remains nearly unchanged and $n_p$ shows a slight reduction ($\sim$ 2\%) at the center. This implies that the alpha-particle abundance is not significantly affected by the presence of coherent structures.

\section{Conclusion and discussion} \label{sec:conclusion and diss}

In this study, we investigate the thermal behavior of alpha particles near 169,141 coherent structures observed by PSP in the inner heliosphere and, to our knowledge, provide the first statistical comparison of their localized response with that of protons. Through analysis of ion plasma parameters centered on small-scale structures selected using different PVI thresholds, we quantify the species-dependent thermal response of the ions within intermittent magnetic structures. Our results reveal three principal features. First, coherent structures are associated with significant ion temperature enhancement, with protons exhibiting a relative larger temperature increase than alpha particles. Second, this differential response produces a localized decrease in the median alpha-to-proton temperature ratio $\langle T_{\alpha}/T_p \rangle$ at the structure center, indicating a reduction in the thermal disequilibrium between the two ion species within these intermittent regions. Third, the localized temperature increase is anisotropic, preferentially enhancing the perpendicular temperature and thereby increasing the $\langle T_{\perp}/T_{\parallel} \rangle$ ratio. Together, these results support the interpretation that coherent structures play an important role in regulating species-dependent and anisotropic ion heating in the young solar wind.

Our results demonstrate that coherent structures are not merely passive features of solar wind turbulence but are regions characterized by strong kinetic effects \citep{2012PhRvL.108d5001S}. In particular, small-scale magnetic flux ropes (SMFRs), which can arise naturally from quasi-two-dimensional MHD turbulence, have been proposed as important sites of particle energization \citep{2015ApJ...814..137Z,2017ApJ...835..147Z,2015ApJ...801..112L,2018ApJ...852L..23Z,2019ApJ...872....4Z,2020ApJ...903...76C}. A “sea” of contracting and merging SMFRs can accelerate particles through first and second-order Fermi processes, reconnection electric fields, and repeated stochastic interactions \citep{2015PhPl...22j0704D,2018ApJ...866....4L}. Parker-type and focused kinetic transport theories incorporating these processes have shown promising results in reproducing power-law-like spectra and energetic ion flux enhancements associated with SMFRs in the inner heliosphere as well as beyond the heliospheric termination shock \citep{2017ApJ...843....4K,2019ApJ...873...72A,2018ApJ...864L..34Z}. Although these theories primarily address energetic-particle acceleration, they further support the broader view that dynamically evolving coherent structures can serve as localized sites of energy conversion.

While $T_{\alpha}/T_{p}$ in the background young solar wind is typically around 5--7, passage through a coherent structure can reduce this disparity by up to 12\% in some events. Since protons have a lower mass-to-charge ratio than alpha particles, they may acquire a larger increase in thermal energy relative to their background state, thereby narrowing the temperature difference between the two species. The positive correlation between $\delta v_{\alpha,p}$ and $T_\alpha/T_p$ at the structure center is consistent with the general trend reported by \citet{2024ApJ...977...27P}, suggesting that preferential heating and acceleration of heavy ions may coexist in the young solar wind \citep{2017ApJ...849..126K}. Additional diagnostics, including velocity distribution functions, fluctuation amplitudes at ion-gyroscales, resonance conditions, stochasticity parameters, and wave activity, are required to clarify the physical mechanisms responsible for these thermal signatures. Candidate mechanisms include stochastic heating \citep{2025PhRvL.135y5201B}, wave--particle interactions \citep{2019PhRvL.122c5102S}, and adiabatic processes associated with velocity shears that can generate temperature anisotropy \citep{2024JPlPh..90f9002O}.

The present analysis uses PSP encounter observations over a heliocentric-distance range of 0.05--0.28 AU. The proton and alpha-particle temperature signatures persist across this range, suggesting that the species-dependent thermal response near coherent structures is a consistent feature of the young solar wind. A dedicated study of its radial variation is left for future work. Whether the observed heating signatures are related to crossings of, or proximity to, the Alfvén critical surface \citep{2025ApJ...991L..35M} warrant further investigation. Our preliminary analysis of Wind observations at 1 AU also reveals similar signatures of alpha-particle heating near coherent structures. A systematic comparison between PSP observations in the young solar wind and Wind observations near 1 AU will be presented in our next study, with the aim of clarifying how species-dependent intermittent heating evolves with heliocentric distance.

\section{acknowledgements} \label{sec:acknowledgements}
We greatly appreciate the PSP development and operations teams, as well as the instrument PIs for data access and support. We thank the reviewer for the constructive comments and suggestions that helped improve the manuscript. TYW was supported the NSFC grant 42104157, and the Xingdian Yingcai funding from Yunnan Province. D.V. is supported by STFC Consolidated Grant ST/W001004/1. LSV was supported by Agenzia Spaziale Italiana (ASI) project “Radial Evolution of large- and Kinetic-scale Processes in the Expanding Solar wind (REKIPES)" C83C25000900005, by the Swedish Research Council (VR) Research Grant N. 2022-03352, and by the Swedish National Space Agency (SNSA) Research Grant N. 2025-00229. LSV was supported by the International Space Science Institute (ISSI) in Bern, through ISSI International Team project \#23-591 (Evolution of Turbulence in the Expanding Solar Wind). The level-2 magnetometer data can be found at \citep{Bale_MacDowal_Koval_Pulupa_Quinn_Schroeder_2020}, the level-3 SPAN proton and alpha data can be found at \citep{Livi_Larson_Rahmati_2025,Livi_Larson_Rahmati_20252}.

\bibliographystyle{aasjournal}
\nocite{*}
\bibliography{classify}
%\vspace{5mm}

\end{document}